\documentclass[aps,prl,groupedaddress,superscriptaddress,twocolumn]{revtex4}

\usepackage{graphicx, amsmath, amssymb, subfigure}
\usepackage{hyperref}
\usepackage{color}
\usepackage{float}
\usepackage{chngcntr}
\usepackage{braket}
\usepackage[dvipsnames]{xcolor}

\begin{document}

\title{Unified model for non-Abelian braiding of Majorana and Dirac fermion zero modes}
\date{\today}
	
\author{Tianyu Huang}
\thanks{These authors equally contribute to this article.}
\affiliation{Interdisciplinary Center for Theoretical Physics and Information Sciences, Fudan University, Shanghai 200433, China}

\author{Rui Zhang}
\thanks{These authors equally contribute to this article.}
\affiliation{International Center for Quantum Materials, School of Physics, Peking University, Beijing 100871, China}

\author{Xiaopeng Li}
\affiliation{State Key Laboratory of Surface Physics and Institute for Nanoelectronic Devices and Quantum Computing, Fudan University, Shanghai 200433, China}
\affiliation{Department of Physics, Fudan University, Shanghai 200433, China}
\affiliation{Shanghai Qi Zhi Institute, AI Tower, Xuhui District, Shanghai 200232, China}

\author{Xiong-Jun Liu}
\thanks{xiongjunliu@pku.edu.cn}
\affiliation{International Center for Quantum Materials, School of Physics, Peking University, Beijing 100871, China}
\affiliation{International Quantum Academy, Shenzhen 518048, China}
\affiliation{Hefei National Laboratory, Hefei 230088, China}

\author{X. C. Xie}
\thanks{xcxie@fudan.edu.cn}
\affiliation{Interdisciplinary Center for Theoretical Physics and Information Sciences, Fudan University, Shanghai 200433, China}
\affiliation{International Center for Quantum Materials, School of Physics, Peking University, Beijing 100871, China}
\affiliation{Hefei National Laboratory, Hefei 230088, China}

\author{Yijia Wu}
\thanks{yijiawu@fudan.edu.cn}
\affiliation{Interdisciplinary Center for Theoretical Physics and Information Sciences, Fudan University, Shanghai 200433, China}
\affiliation{State Key Laboratory of Surface Physics and Institute for Nanoelectronic Devices and Quantum Computing, Fudan University, Shanghai 200433, China}
\affiliation{Hefei National Laboratory, Hefei 230088, China}

\begin{abstract}
Majorana zero modes (MZMs) are the most intensively studied non-Abelian anyons. The Dirac fermion zero modes in topological insulators, which are symmetry-protected {\em doubling} of MZMs under fermion number conservation, offer an alternative approach to explore non-Abelian anyons. However, a unified model that elucidates the braiding statistics % intrinsic connection between the quantum statistics
of these types of topological zero modes remains absent. We show that the minimal Kitaev chain model beyond fine-tuning regime provides a unified characterization of the non-Abelian statistics of both MZMs and Dirac fermion zero modes in different parameter regimes. In particular, we introduce a minimal tri-junction setting based on the minimal Kitaev chain model and show it facilitates the unified scheme of braiding Dirac fermion zero modes, as well as the MZMs in the assistance of a Dirac mode. This unified minimal model provides deeper insights into non-Abelian statistics, demonstrating that the non-Abelian braiding of MZMs can be continuously extended to encompass Dirac fermion zero modes.  The minimal Kitaev chain has been realized in coupled quantum dots [\textit{Nature} \textbf{614}, 445 (2023)]. Our extension, which demonstrates novel nontrivial phases with non-Abelian MZM pairs and Dirac zero modes emerging in the broader parameter regimes without fine-tuning, expands the accessible experimental parameter space and enhances the feasibility of observing non-Abelian statistics in the minimal Kitaev chain model.
%\textbf{Keywords:} anyons, geometric and topological phases, topological quantum computing, topological insulators, topological superconductors
\end{abstract}
	
\maketitle
	
\textit{Introduction.}
Anyons~\cite{Wilczek1982} in condensed matter systems usually emerge as composites formed from charged particles~(like electrons) and vortices. This anyon model was raised by Wilczek~\cite{Wilczek1982} and first exemplified by the Laughlin quasiparticles~\cite{Laughlin1983} in the fractional quantum Hall (FQH) system~\cite{FQH1982}.
%Laughlin quasiparticles obeying Abelian anyons statistics can be regarded as elementary excitations in composite fermion~\cite{} systems in which electrons are bound with integer flux.
Also in the $\nu=5/2$ FQH state~\cite{WillettFQH5_2}, the more intriguing non-Abelian anyons are proposed to emerge as the elementary excitations binding with half-vortices~\cite{MooreRead1991, XGWen1991}. This type of excitation also manifests itself as a half-vortex-bound state in $p$-wave superconductors~\cite{Ivanov_2001}, which is well known as the Majorana zero mode~(MZM)~\cite{KitaevChain_2001}. Due to their significant potential for fault-tolerant topological quantum computation~\cite{Kitaev2003TQC, RMP_TQC}, substantial efforts have been dedicated to seeking experimental evidence of MZMs in various platforms including superconductor-semiconductor heterostructures~\cite{Fu_and_Kane, kou, das1, DELiu_review}, ferromagnetic atomic chains on a superconductor~\cite{perge}, and vortices in topological superconductors~\cite{Jia, Fes1, Fes3, HDing_review}. However, the complexity of the superconductor-proximitized structure~\cite{XGWen_CMEM, DELiu_review} and the possible trivial explanations for the transport signals~\cite{JieLiu_ZBP_2012_PRL, DasSamra_ABSvsMBS} still hinder the appearance of the smoking-gun experimental signals.

An alternative approach for non-Abelian anyons is the symmetry-protected Majorana pairs. The most studied one among them is the Majorana Kramer's pair (MKP)~\cite{KTLaw_MKP, Mele_MKP}, where a pair of MZMs is related by time-reversal (TR) symmetry. Owing to the TR symmetry protection, MKPs have been shown to obey a novel type of symmetry-protected non-Abelian statistics~\cite{XJLiu_MKP, Flenseberg_MKP_braiding_1, Flenseberg_MKP_braiding_2, XJLiu_MKP_braiding, Fu_MKP_TQC}. When the TR symmetry~(an anti-unitary symmetry) in the MKP is replaced by a unitary symmetry, this type of Majorana pair will undergo a continuous crossover to a zero-energy Dirac fermion mode bound to a half vortex~\cite{Wu_HOTI, yasui2012dirac, hong2022unitary} when fermion number conservation is imposed~\cite{hong2022unitary, hong2024parafermions}. Here, the attached half vortex is regarded as the origin of the non-Abelian statistics of these Dirac fermion zero modes~\cite{yasui2012dirac}.
%A two dimensional disk with a half-vortex penetrated can be continuously deformed into a
This kind of zero-energy Dirac fermion mode can appear as a bound state with a half-vortex attached in a two-dimensional topological insulator~\cite{SQShen_PRB_2011}, or as the celebrated Jackiw-Rebbi zero mode in one-dimensional system, whose $1/2$ fractional charge~\cite{JackiwRebbi, SSH} and fractional statistics~\cite{DanileLoss_2013_PRL, DanileLoss_2015_PRB, SSH_braiding} were discussed in early studies.

Although the braiding statistics of the Dirac fermion zero modes has an intrinsic relation to MZM pairs~\cite{hong2022unitary}, a unified model that simultaneously realizes non-Abelian statistics of MZMs and Dirac fermion zero modes is lacking. %More specifically,
Currently, the three distinct classes of non-Abelian braiding statistics have been only studied separately, namely, respectively for MZMs without the need of symmetry protection, whose braiding can be assisted by quantum-dot~(QD)~\cite{flensberg2011non, liu2021minimal, votexMZM_QD_assisted, Flensberg_QD_assisted, xu2023dynamics}, %in which two MZMs are braided through coupling to a single Dirac fermion zero mode,
the symmetry protected Majorana pairs, and Dirac fermion zero modes. %The latter two classes necessitate symmetry protection,
%Unlike Majorana modes in topological superconductors, the Dirac fermion zero modes can be obtained in topological insulators with particle number conservation. %former is obtained when assigning the fermion number conservation to the latter as an intermediary framework bridging MZMs and Dirac fermion zero modes.
Nevertheless, whether the three classes of braiding statistics can be unified in a single quantum model was not known. Such a unified model not only is fundamentally significant, but also may show insights into the ultimate experimental realization of the non-Abelian statistics. % Despite their interconnections, a unified description covering these three distinct but interconnected schemes is still lacking.}
%In addition, it has long been discussed that the braiding operation of MZMs can be facilitated by a Dirac fermion zero mode. In this braiding scheme, two MZMs are braided through coupling to a single Dirac fermion zero mode, significantly simplifying the braiding operation. Consequently, the low-lying states involved include both the MZMs and Dirac fermion modes, making QD-assisted Majorana braiding a bridging scheme between Majorana and Dirac fermionic systems.

In this article, we show that the minimal Kitaev chain~\cite{PMM_2012, PMMs_PRXquantum} can play the role of such a unified model via constructing proper Y-junction configuration. In particular,
when the electron hopping amplitude coincides with the superconducting pairing, the minimal Kitaev chain based Y-junction captures the non-Abelian braiding of MZMs (including the QD-assisted Majorana braiding). Conversely, when the electron hopping and superconducting pairing are different, the model can realize the non-Abelian braiding of Majorana pairs. Notably, when the superconducting pairing amplitude continuously decreases to zero, the non-Abelian Majorana pairs undergo a continuous crossover to non-Abelian Dirac fermion zero modes. %This indicates that the quasiparticles being braided in our setting can switch from the MZMs to Dirac fermion zero modes with a small change in parameters.
This result shows that with this unified model the non-Abelian braiding can be realized over a broad experimental parameter regime without fine tuning. Particularly, the non-Abelian braiding statistics of the Dirac zero modes can be achieved by relaxing the constraint on fine tuning of the electron hopping and superconducting pairing in the minimal Kitaev chain based Y-junction. Our finding highlights that non-trivial physics can also emerge within this newly extended parameter regime, which is of high experimental feasibility and can be implemented using ultracold atoms~\cite{Yb_ColdAtom} or QD arrays~\cite{MKC_exp_1, MKC_exp_2, MKC_exp_3, threeSite_KitaevChain_Kouwenhoven, threeSite_KitaevChain, probeMajorana_KitaevChain_Kouwenhoven}. The latter (QD array) is closely related to the very recent experimental advancements \cite{threeSite_KitaevChain_Kouwenhoven, threeSite_KitaevChain, probeMajorana_KitaevChain_Kouwenhoven}.

%%%%%%%%%%%%%%%%%%%% I AM THE CUTTING LINE %%%%%%%%%%%%%%%%%%%%

\textit{Braiding in minimal Kitaev chain.}
The celebrated Kitaev's chain~\cite{KitaevChain_2001} model %describes one-dimensional spinless $p$-wave superconductor supporting
supports MZMs as its topological edge states. The two-site version of the Kitaev's chain is known as the minimal Kitaev chain~\cite{PMM_2012, PMMs_PRXquantum}, which can be described by \(H = \chi d_L^\dagger d_R + \Omega d_L^\dagger d_R^\dagger + \text{h.c.}\). Here, the index $L$~($R$) denotes the left~(right) site. %It is worth noting that, d
Despite lacking a continuous $k$-space, the minimal Kitaev chain can still be depicted by a $\mathbb{Z}_2$ topological index~\cite{KitaevChain_2001, Z2_index_DasSarma_1, Z2_index_DasSarma_2, Z2_index_SQShen} through the Pfaffian of the Hamiltonian %in the Majorana basis
at momentum $k=0$ and $k=\pi$. The minimal Kitaev chain reaches the ``sweet spot''~\cite{PMMs_PRXquantum, Chun-Xiao_tuning} when the hopping amplitude $\chi$ is equal to the superconducting pairing amplitude $\Omega$ and both of them are real numbers (\(\chi = \Omega \in \mathbb{R}\)), where the Hamiltonian above reduces to \(H = -i|\chi| \cdot i(d_L^\dagger - d_L)(d_R + d_R^\dagger)\). In this way, two isolated MZMs appear at the left and the right site as \((d_L + d_L^\dagger)\) and \(i(d_R^\dagger - d_R)\), respectively, which are termed poorman's MZMs~(PMMs)~\cite{PMM_2012, PMMs_PRXquantum}.
%Despite the absence of topological protection, PMMs still exhibit non-Abelian braiding properties identical to those of conventional MZMs.

\begin{figure}[t]
(a) \includegraphics[width=0.16\textwidth]{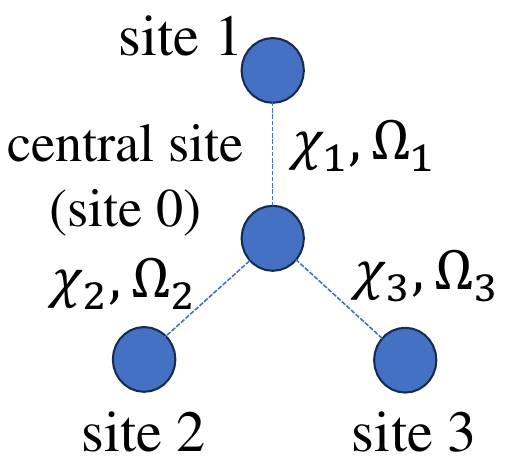}
(b) \includegraphics[width=0.2\textwidth]{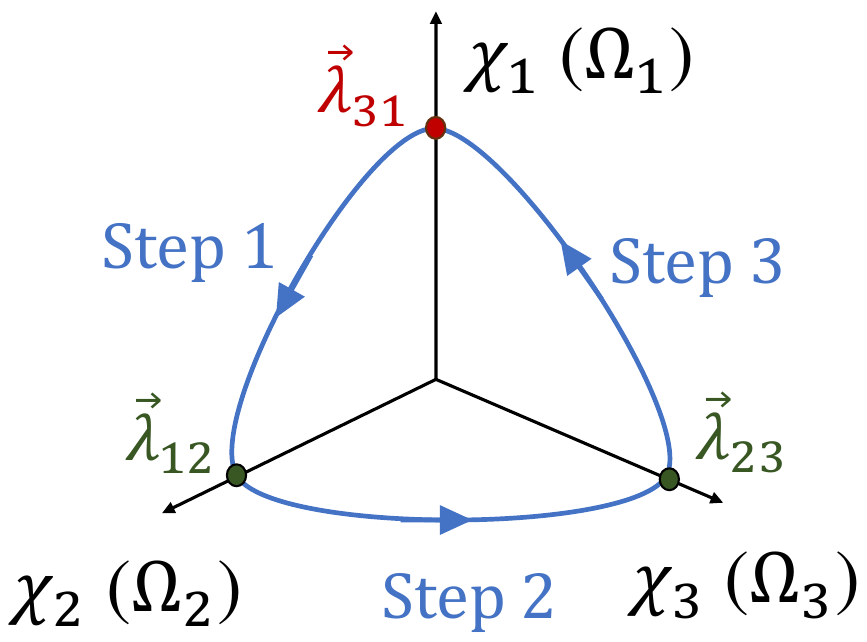}
\caption{(a) Sketch of the Y-junction jointed by minimal Kitaev chains. (b) The braiding operation corresponds to a closed loop (started and ended at the point marked in red) in the parameter space.}
\label{device}
\end{figure}

However, properly defining non-Abelian braiding requests at least four MZMs~\cite{Ivanov_2001}. Therefore, in order to perform the braiding of the PMMs, the minimal device required is a four-site Y-shaped junction~\cite{Fisher_braiding_2011, JDSau_braiding_2011, von2017topological, Yjunction_Trif_Simon} jointed by three minimal Kitaev chains [see Fig. \ref{device}(a)]
\begin{eqnarray}
    H = \sum\limits_{j=1}^{3} \chi_j d_0^{\dagger} d_j - \chi_j^* d_0 d_j^{\dagger} +\Omega_j d_0^{\dagger} d_j^{\dagger} - \Omega_j^* d_0 d_j.
\label{Yjunction_Hamiltonian}
\end{eqnarray}
Here, $d_j$ ($d_j^{\dagger}$) annihilates (creates) an electron on site $j$, %located at the outer ends
while $d_0$ ($d_0^{\dagger}$) annihilates (creates) an electron on the central site, $\chi_j$ and $\Omega_j$ are the amplitudes of the hopping and superconducting pairing, respectively. In general, $\chi$'s and $\Omega$'s are complex as $\chi_j = |\chi_j| e^{i \theta_{\chi_j}}$ and $\Omega_j = |\Omega_j| e^{i \theta_{\Omega_j}}$.
Significantly, the ``sweet spot'' condition is satisfied when $|\chi_j|=|\Omega_j|$, and $\theta_{\chi_j}, \theta_{\Omega_j} = 0$, which requires fining tuning of the experimental parameters.
%Significantly, it is impossible to gauge away all these six phases $\theta_{\chi_j}$ and $\theta_{\Omega_j}$ ($j=1,2,3$), since in Eq. (\ref{Yjunction_Hamiltonian}), there are only four electron operators can be utilized to absorb the gauge phases. In other words, the ``sweet spot'' condition cannot always be satisfied when we consider the braiding in this minimal-Kitaev-chain-based Y-junction.
Nonetheless, the non-Abelian braiding physics shall not collapse due to a slight deviation of the parameters from the ``sweet spot''. Meanwhile, the low-lying states of this Y-junction being braided now are not necessarily PMMs. %when one deviates from this ``sweet spot''.
It implies that this Y-junction can depict the non-Abelian braiding of different kinds of low-lying modes for different choices of $\chi$'s and $\Omega$'s. Specifically, there are four distinct cases based on the magnitudes and the phases of $\chi_j$ and $\Omega_j$, which are summarized in TABLE \ref{table}. The fact that the proposed Y-junction can realize the non-Abelian braiding of Majorana pairs and Dirac fermion zero modes with control, in addition to the single MZMs shall significantly mitigate experimental challenges in realizing the non-Abelian statistics, as studied below in detail. %In this manner, the non-Abelian anyon braided at this Y-junction extends beyond the realm of MZMs. Consequently, the experimental challenges are significantly mitigated, as the constraints on parameters are substantially alleviated.}

%Specifically, the parameters $\chi_j$ and $\Omega_j$ are classified according to the relationships between their magnitudes and phase arguments, with discussions of the associated braiding modes in each case summarized in TABLE~\ref{table 1}.

\begin{table}[t]
\centering
\addtolength{\tabcolsep}{1.5pt}
\begin{tabular}{c p{3.3cm} p{3.1cm}}
\hline
\hline
 & $\theta_{\chi_j}, \theta_{\Omega_j} = 0$ & $\theta_{\chi_j}, \theta_{\Omega_j} \neq 0$ \\ \hline
 $|\chi_j| = |\Omega_j|$ &Case \uppercase\expandafter{\romannumeral 1}: poorman's MZMs (PMMs) & Case \uppercase\expandafter{\romannumeral 2}: QD-assisted MZM braiding \\ \hline
 $|\chi_j| \neq |\Omega_j|$ &Case \uppercase\expandafter{\romannumeral 3}: unitary-symmetry protected Majorana pairs/Dirac fermion zero modes & Case \uppercase\expandafter{\romannumeral 4}: unitary-symmetry broken Majorana pairs/Dirac fermion zero modes \\
 \hline
 \hline
\end{tabular}
\addtolength{\tabcolsep}{-1.5pt}
\caption{The minimal-Kitaev-chain-based Y-junction can depict the non-Abelian braiding of different types of low-lying modes depending on the choices of $\chi$'s and $\Omega$'s.}
\label{table}
\end{table}

\begin{figure*}[t]
\includegraphics[width=0.24\textwidth]{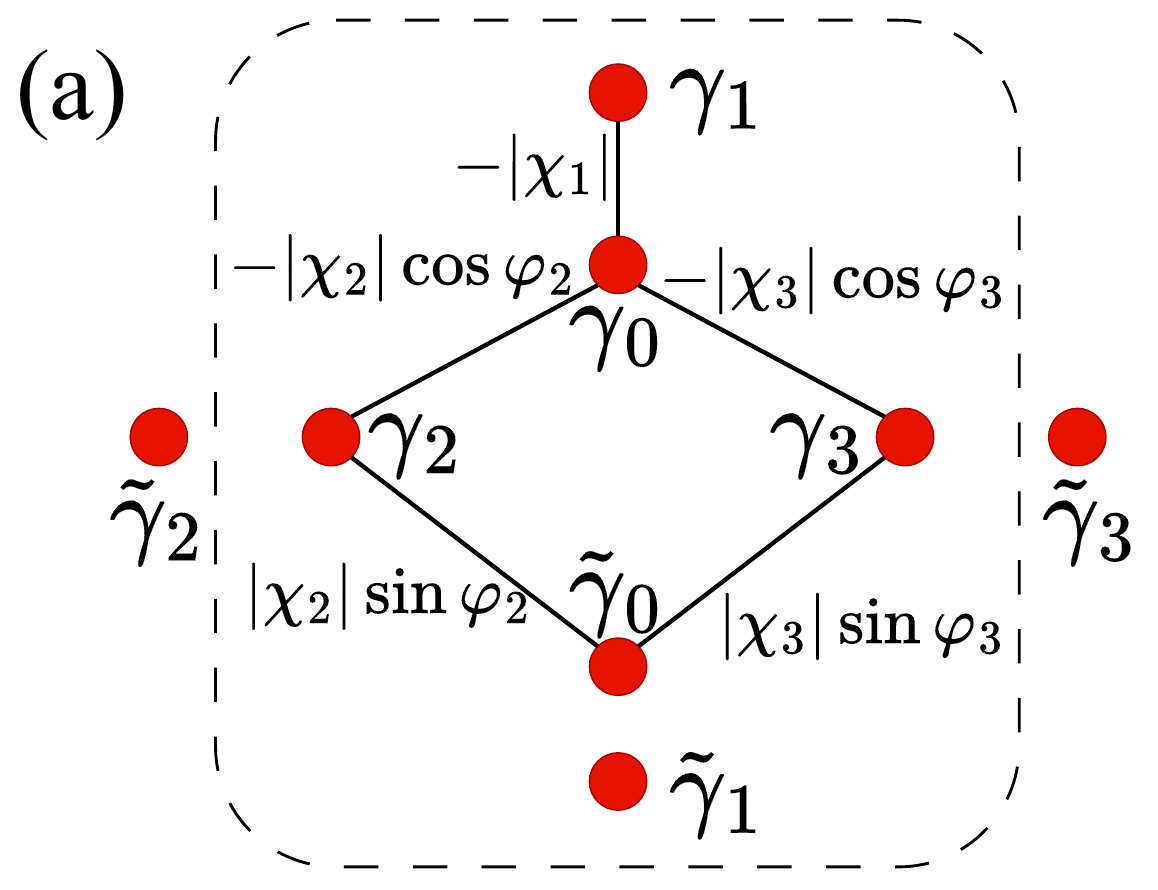}
\includegraphics[width=0.24\textwidth]{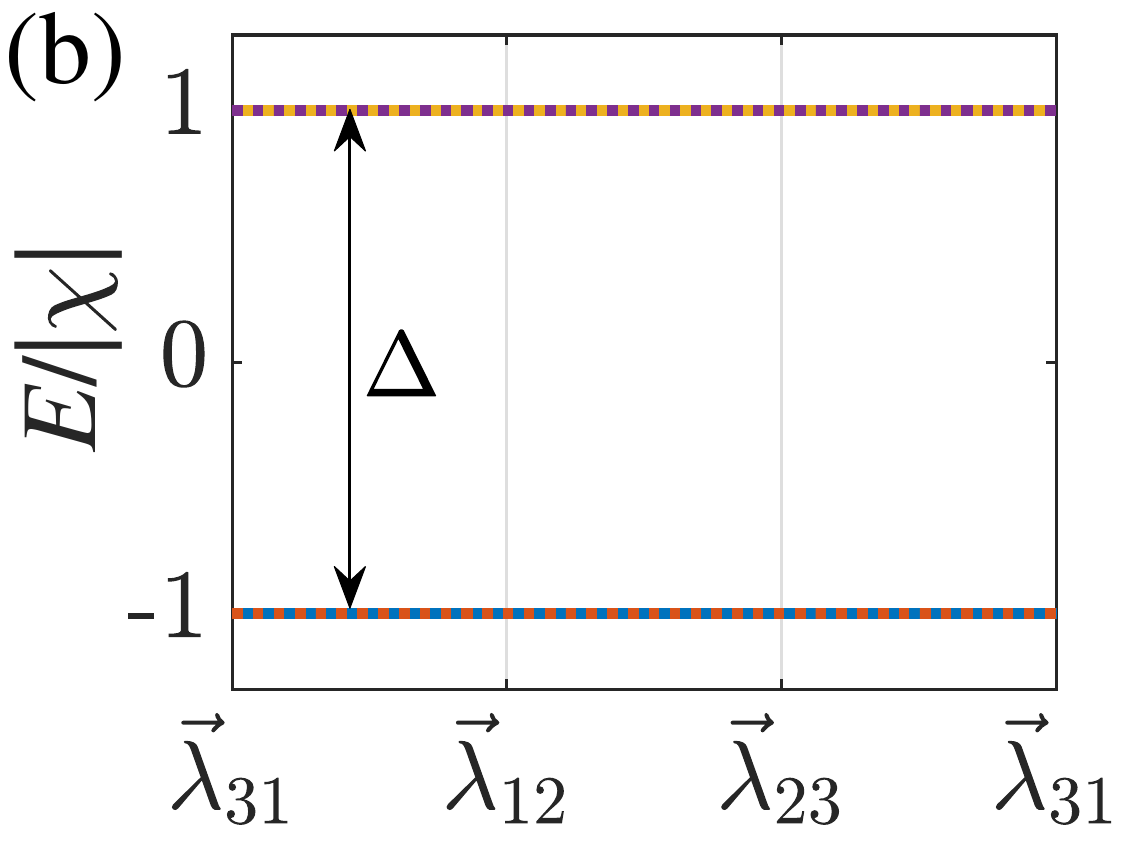}
\includegraphics[width=0.24\textwidth]{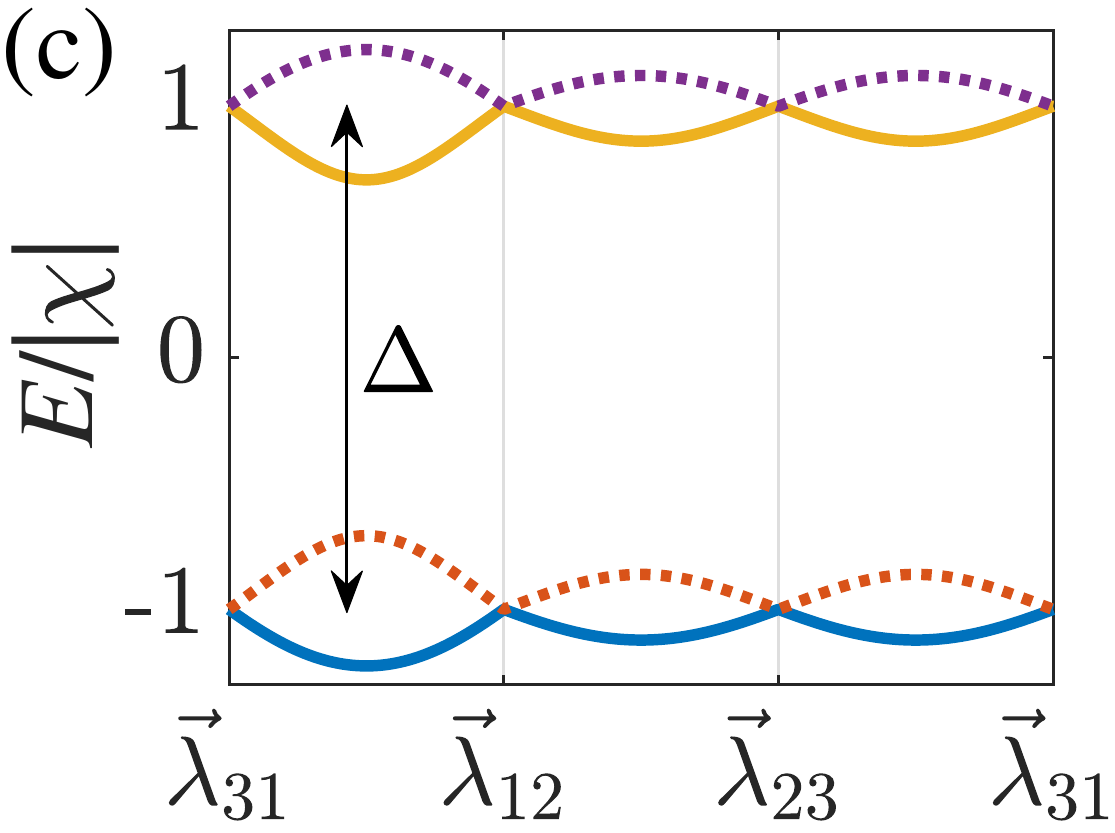}
\caption{(a) Schematic representation of the Majorana Hamiltonian [Eq.~(\ref{Hamiltonian_CaseIandII})] under the conditions that $|\chi_j|=|\Omega_j|$. The Fock space we consider is spanned by six MZMs inside the dashed box.
(b), (c) The energy spectra during the braiding as (b)
Case \uppercase\expandafter{\romannumeral 1} in TABLE \ref{table} that $|\chi_j|=|\Omega_j|$ and $\varphi_2,\varphi_3=0$. All the states in (b) are doubly degenerate; (c) Case \uppercase\expandafter{\romannumeral 2} in TABLE \ref{table} that $|\chi_j|=|\Omega_j|$ and $\varphi_2,\varphi_3\ne0$ ($\varphi_2 = \pi/6$, and $\varphi_3= \pi/12$ here for illustration). $|\chi| \equiv \sqrt{|\chi_1|^2 + |\chi_2|^2 + |\chi_3|^2} $ is chosen as the energy unit.}
\label{CaseIandII}
\end{figure*}

The braiding operation in this Y-junction is conducted through coupling the three outer sites with the central site in turn~\cite{JDSau_braiding_2011, von2017topological}. Without loss of generality, we choose swapping the low-lying modes localized at site 2 and site 3 of this Y-junction [see Fig.~\ref{device}(a)] as an instance. %of the braiding operation
The Y-junction is initialized by setting $\chi_2$, $\chi_3$, $\Omega_2$, and $\Omega_3$ to zero, while setting $\chi_1$ and $\Omega_1$ to finite values. In this way, both site 2 and site 3 become isolated so that each of them contains a zero-energy mode. In the first step of the braiding, $\chi_2$ and $\Omega_2$ are adiabatically increased to finite values, while $\chi_1$ and $\Omega_1$ are adiabatically decreased to zero. Here, we assume the ratios $|\chi_1|/|\Omega_1|$ and $|\chi_2|/|\Omega_2|$ remain constants during the braiding, and the phases $\theta_{\chi_j}, \theta_{\Omega_j}$ also remain unchanged. In this way, the zero-energy mode initially localized at site 2 is adiabatically moved to site 1. Similarly, in the second step, $\chi_3$ and $\Omega_3$ are both increased to finite values, and $\chi_2$ and $\Omega_2$ are decreased to zero in an adiabatic way so that the zero-energy mode in site 3 is moved to site 2. In the final step, the braiding is accomplished by increasing $\chi_1$ and $\Omega_1$ to their initial values while decreasing $\chi_3$ and $\Omega_3$ to zero. Such a braiding process corresponds to a closed loop in the parameter space [see Fig. \ref{device}(b)].

%%%%%%%%%%%%%%%%%%%% I AM THE CUTTING LINE %%%%%%%%%%%%%%%%%%%%

\textit{Non-Abelian braiding of MZMs.}
We firstly focus on the braiding under the conditions $|\chi_j|=|\Omega_j|$~($j=1,2,3$), i.e. case \uppercase\expandafter{\romannumeral 1} and case \uppercase\expandafter{\romannumeral 2} in TABLE~\ref{table}. %and analytically solve for the non-Abelian Berry phase.
By decomposing the electron operators $d_j$ into Majorana ones $\gamma_j$ and $\widetilde{\gamma}_j$, the original Hamiltonian [Eq.~(\ref{Yjunction_Hamiltonian})] can be rewritten as~\cite{SupplementaryMaterials}:
\begin{eqnarray}
%\begin{split}
    H &= &-i|\chi_1| \gamma_0 \gamma_1 -i|\chi_2|( -\sin \varphi_2 \widetilde{\gamma} _0 + \cos \varphi_2 \gamma_0 )\gamma_2\nonumber\\
     &&-i|\chi_3|( -\sin \varphi_3 \widetilde{\gamma}_0 + \cos \varphi_3 \gamma_0 )\gamma_3.
%\end{split}
\label{Hamiltonian_CaseIandII}
\end{eqnarray}

\noindent Here, %as we have stated before,
$\varphi_2 \equiv (\theta_{\chi_2} + \theta_{\Omega_2} - \theta_{\chi_1} - \theta_{\Omega_1})/2 $ and $\varphi_3 \equiv (\theta_{\chi_3} + \theta_{\Omega_3} - \theta_{\chi_1} - \theta_{\Omega_1})/2$ are two phase constants~($\theta_{\chi_j}$ and $\theta_{\Omega_j}$ remain unchanged during the braiding) that cannot be absorbed even after gauge transformation~\cite{SupplementaryMaterials}. This Majorana Hamiltonian [Eq.~(\ref{Hamiltonian_CaseIandII})] can be schematically presented by Fig.~\ref{CaseIandII}(a). Specifically, we consider the %$\sqrt{2}^6/2$-dimensional
parity-even subspace spanned by six MZMs $\gamma_0$, $\gamma_1$, $\gamma_2$, $\gamma_3$, $\widetilde{\gamma}_0$, and $\widetilde{\gamma}_1$ [dashed box in Fig.~\ref{CaseIandII}(a)], because fermion parity is conserved here and the Hamiltonian in the parity-even subspace is identical to that in the parity-odd subspace~(up to a constant)~\cite{SupplementaryMaterials}. %It is worth noting that
Although $\widetilde{\gamma}_1$ is isolated from the other MZMs, it is still intentionally considered to ensure that the Fock space dimension is an integer.

%Furthermore, we emphasize that the global degeneracy of the energy spectrum plays a crucial role.

In the case \uppercase\expandafter{\romannumeral 1} of TABLE~\ref{table} that $\varphi_2, \varphi_3 = 0$, the two-fold degenerate negative-energy states are gapped from the two-fold degenerate positive-energy states (the gap is denoted by $\Delta$), and these degeneracies are preserved throughout the braiding [see Fig.~\ref{CaseIandII}(b)]. In the adiabatic limit, where the braiding time $T \gg 1/\Delta$, it is meaningful to consider only the two-fold degenerate ground state manifold. The Berry phase during the braiding can be evaluated analytically~\cite{SupplementaryMaterials}, which leads to an evolution operator as $U_{\mathrm{case~I}} = \begin{pmatrix}
    e^{-i\pi/4} & 0 \\
        0 & e^{i\pi/4}
\end{pmatrix}$. This is precisely the Majorana braiding operator~\cite{Ivanov_2001} as expected, since the ``sweet spot'' condition is satisfied in this case, ensuring that all the low-lying modes here are PMMs.

In the case \uppercase\expandafter{\romannumeral 2} of TABLE~\ref{table} that $\varphi_2, \varphi_3 \ne 0$, the two-fold ground-state degeneracy is lifted except for three special parameter points during the braiding [see Fig.~\ref{CaseIandII}(c)]. Therefore, in addition to the Berry phase, the dynamic phase in the evolution operator $U = \hat{P} \exp[\int -iE(t) \mathrm{d}t + i\mathbf{A}(\boldsymbol{\lambda}) \cdot \mathrm{d}\boldsymbol{\lambda}]$~\cite{BerryRotationMatrix_Wilczek_Zee, chang2021lecture} generally cannot be dropped. Here, $\hat{P}$ represents the time-ordering operator, $E$ is a diagonal matrix corresponding to the eigenenergy spectra, and $\mathbf{A}$ is the Berry connection. Under the adiabatic condition %the braiding time cost
$T \gg 1/\Delta$, we could still focus only on the two lowest states with negative energies [see Fig. \ref{CaseIandII}(c)]. The corresponding evolution operator can be evaluated numerically as $U_{\mathrm{case~II}} = e^{i\theta} \begin{pmatrix}
    e^{-i(\pi/4+\delta)} & 0 \\
        0 & e^{i(\pi/4+\delta)}
\end{pmatrix}$. Here, the average energy of these two lowest states contributes a trivial overall dynamic phase $\theta$, while the energy difference between these two lowest states leads to a dynamic phase difference $\delta$
\cite{SupplementaryMaterials}.
If %$1/w \gg T \gg 1/\Delta$
$\frac{1}{\varphi_2}, \frac{1}{\varphi_3} \gg T \gg \frac{1}{\Delta}$,
%where %$w \sim \max\left\{\sin(\varphi_2/2), \sin(\varphi_3/2)\right\}$
%$w$ indicates the degeneracy splitting scale [see Fig. \ref{CaseIandII}(c)]
then this dynamic phase difference $\delta$ is relatively small so now the braiding operator comes back to the form of $\mathrm{diag}\{e^{-i\pi/4}, e^{i\pi/4}\}$. It indicates that the geometric phase accumulated remains $\mp \pi/4$ even when $\varphi_2, \varphi_3 \ne 0$, which can be confirmed by analytically evaluating the geometric phase part of the evolution operator $\hat{P} \exp[ \int i\mathbf{A}(\boldsymbol{\lambda}) \cdot \mathrm{d}\boldsymbol{\lambda}]$~\cite{SupplementaryMaterials}. Otherwise, the dynamic phase dominates when $\frac{1}{\varphi_2}, \frac{1}{\varphi_3} \ll T $, causing the braiding results to oscillate with increasing $T$~\cite{SupplementaryMaterials}. Such results that in addition to the geometric phase of $\mp \pi/4$, an undesired dynamic phase being also involved has been observed in the QD-assisted braiding~\cite{flensberg2011non, liu2021minimal, votexMZM_QD_assisted, xu2023dynamics}. In fact, as shown in Fig.~\ref{CaseIandII}(a), two MZMs $\gamma_2$ and $\gamma_3$ being braided here are swapped via coupling to both the two Majorana components $\gamma_0$ and $\widetilde{\gamma}_0$ located at the central site. Therefore, the minimal Kitaev chain here exactly describes the QD-assisted braiding that two MZMs are swapped via a Dirac fermion zero mode.

%%%%%%%%%%%%%%%%%%%% I AM THE CUTTING LINE %%%%%%%%%%%%%%%%%%%%

\begin{figure}[t]
\includegraphics[width=0.205\textwidth]{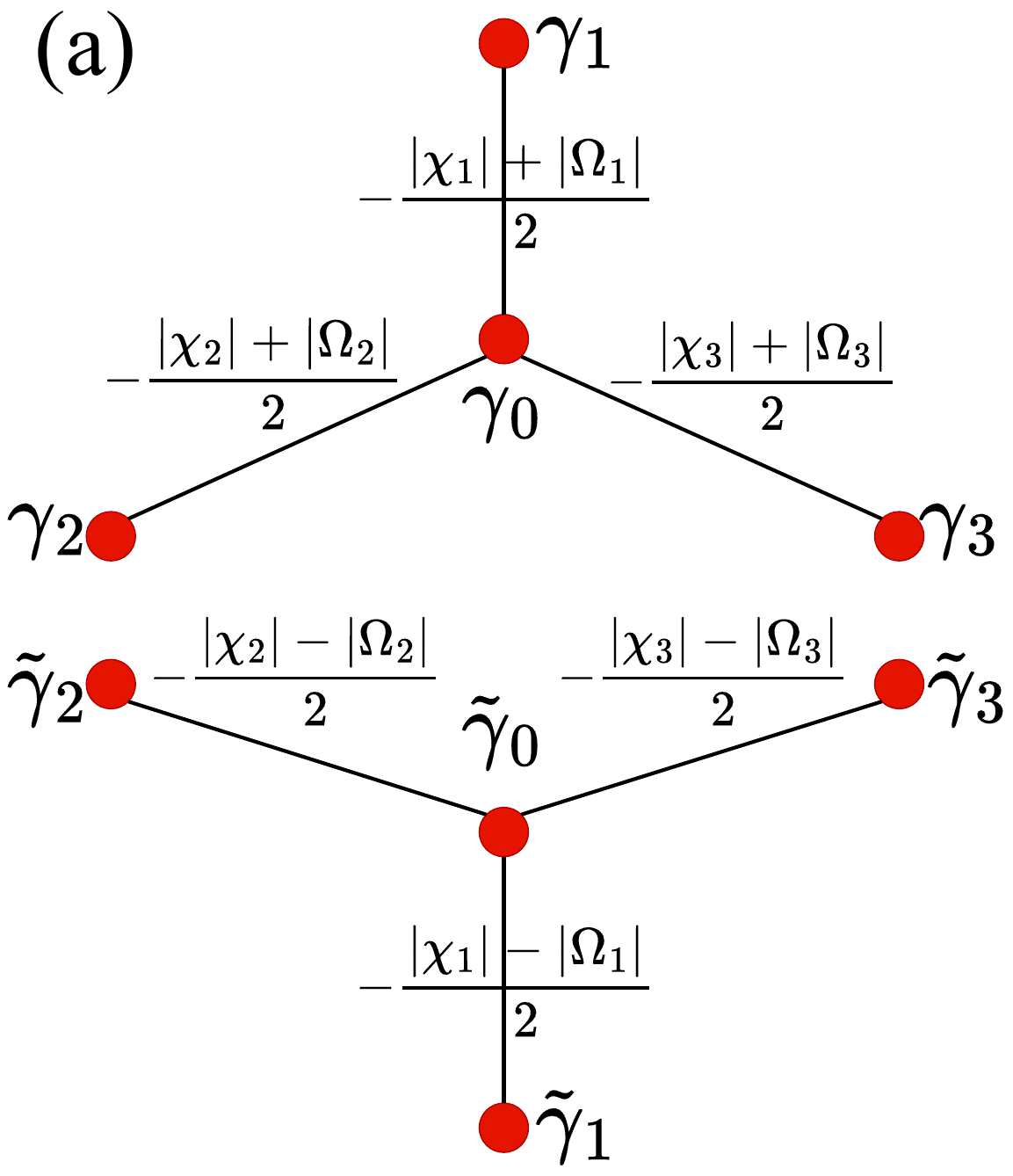}
\includegraphics[width=0.225\textwidth]{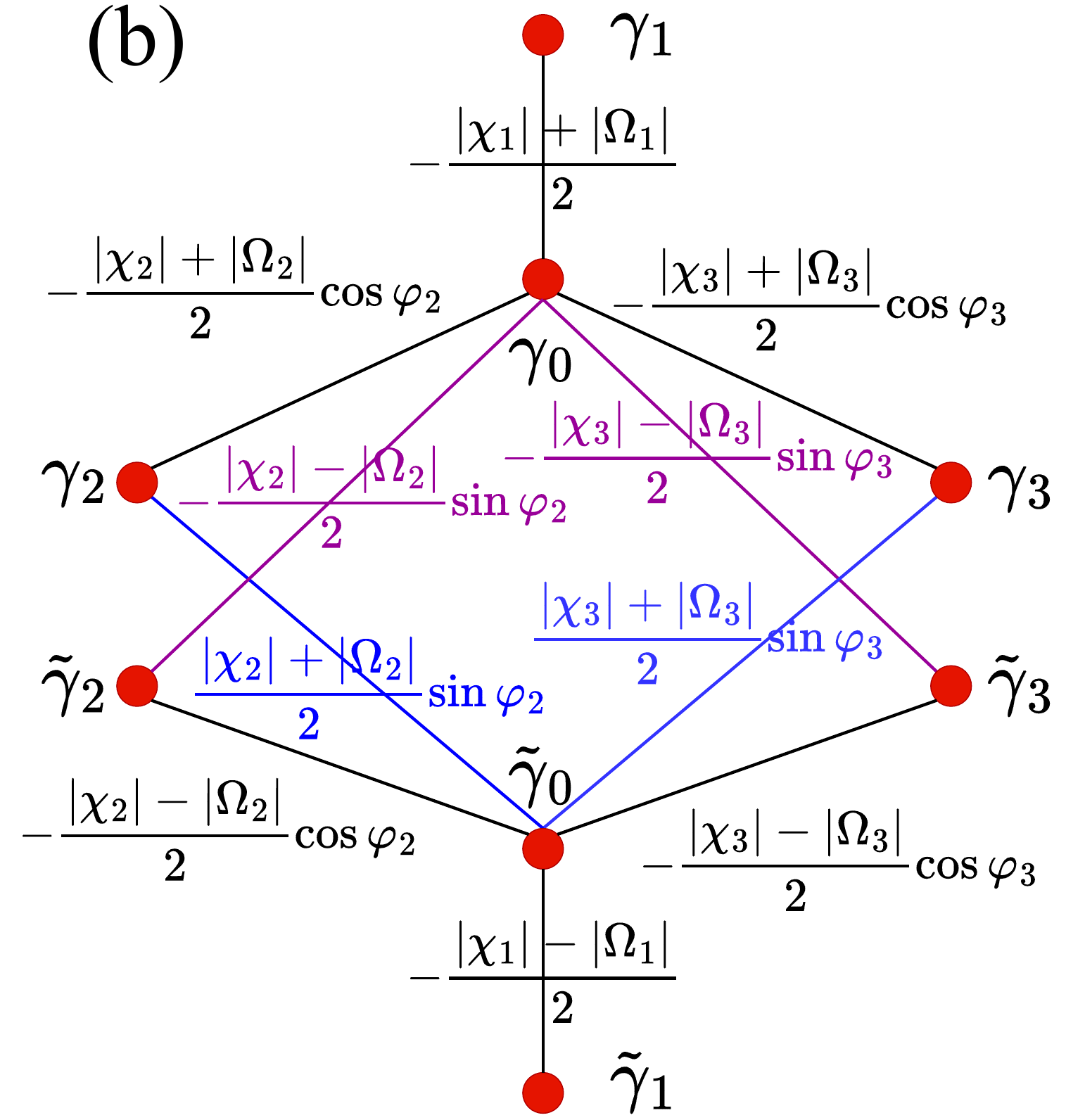}
\includegraphics[width=0.21\textwidth]{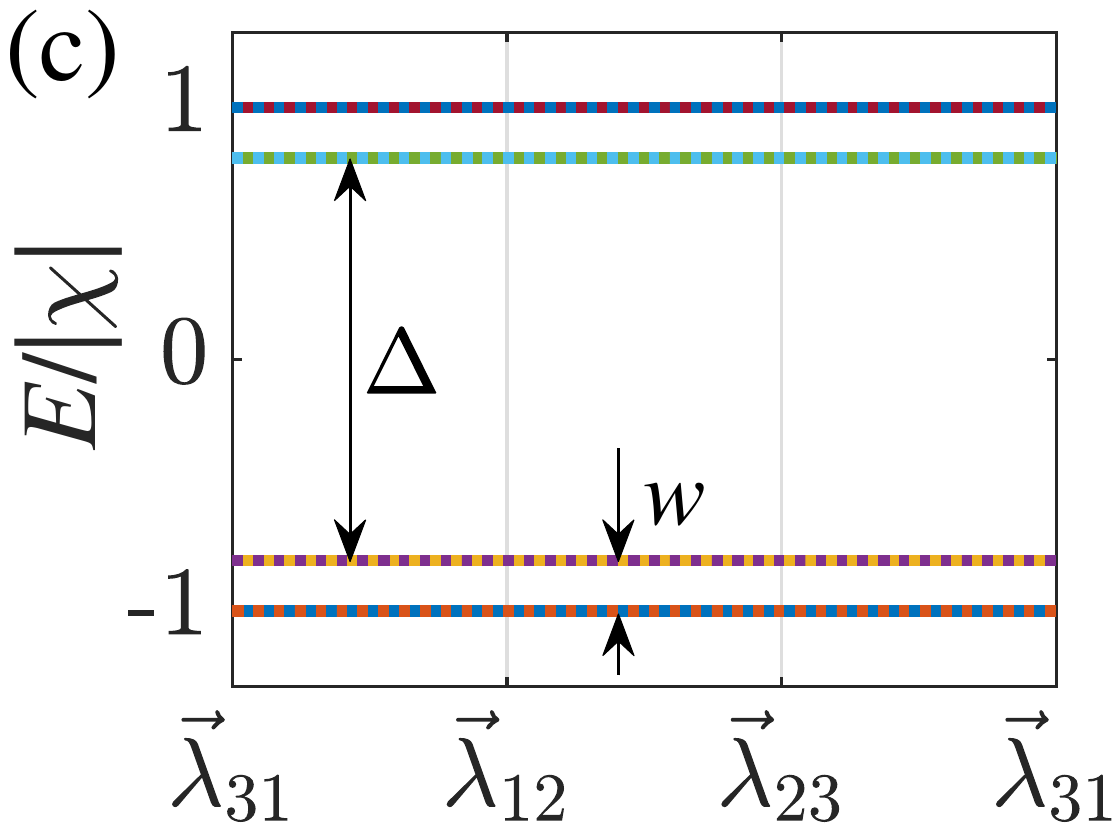}
\includegraphics[width=0.21\textwidth]{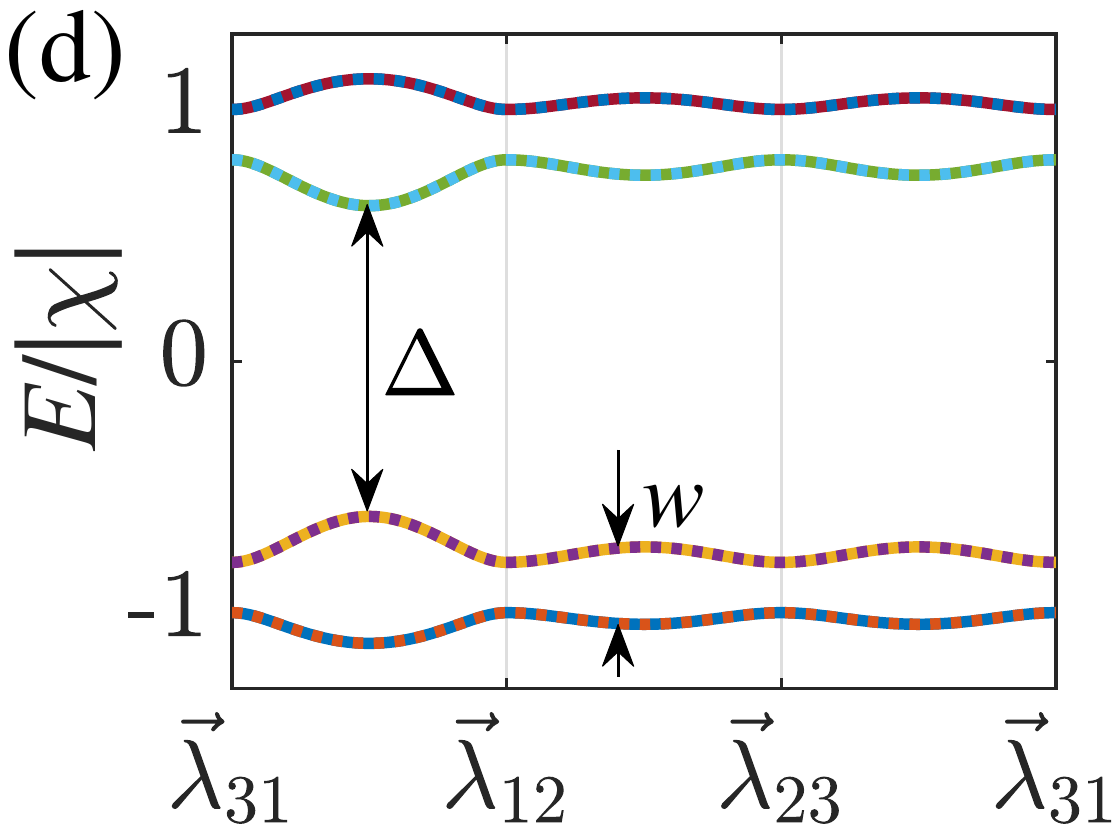}
\caption{(a), (b) Schematic representation of the Majorana Hamiltonian [Eq. (\ref{Hamiltonian_CaseIII})] under the conditions of (a) Case \uppercase\expandafter{\romannumeral 3} in TABLE \ref{table}; and (b) Case \uppercase\expandafter{\romannumeral 4} in TABLE \ref{table}.
(c), (d) The energy spectra during the braiding as (c) Case \uppercase\expandafter{\romannumeral 3} that $|\chi_j| \ne |\Omega_j|$ and $\varphi_2,\varphi_3=0$ ($|\Omega_i|/|\chi_i|=0.8$ here for illustration); and (d) Case \uppercase\expandafter{\romannumeral 4} that $|\chi_j| \ne |\Omega_j|$ and $\varphi_2,\varphi_3 \ne 0$ ($|\Omega_i|/|\chi_i|=0.8$, $\varphi_2 = \pi/6$, and $\varphi_3= \pi/12$ here for illustration). The energy unit $|\chi| \equiv \sqrt{|\chi_1|^2 + |\chi_2|^2 + |\chi_3|^2} $, and all the states in (c) and (d) are doubly degenerate.}
\label{CaseIIIandIV}
\end{figure}

\textit{Non-Abelian braiding of Majorana pairs and Dirac fermion zero modes.}
The more intriguing scenario is that even though the equalities $|\chi_j|=|\Omega_j|$ ($j=1,2,3$) are only slightly violated (an unavoidable circumstance in realistic condition), the low-lying modes being braided are no longer single MZMs, but Majorana pairs. To be specific, under the conditions $|\chi_j|\ne|\Omega_j|$ ($j=1,2,3$), i.e. cases \uppercase\expandafter{\romannumeral 3} and \uppercase\expandafter{\romannumeral 4} in TABLE \ref{table}, the original Hamiltonian [Eq.~(\ref{Yjunction_Hamiltonian})] in the Majorana form reads~\cite{SupplementaryMaterials}
%\begin{align}
%\begin{split}
%H= i \frac{|\chi_1| + |\Omega_1|}{2} \gamma_0 \gamma_1
%+ i \frac{|\chi_2| + |\Omega_2|}{2} \gamma_0 \gamma_2
%+ i \frac{|\chi_3| + |\Omega_3|}{2} \gamma_0 \gamma_3\\
%- i \frac{|\chi_1| - |\Omega_1|}{2} \widetilde{\gamma}_0 \widetilde{\gamma}_1
%- i \frac{|\chi_2| - |\Omega_2|}{2} \widetilde{\gamma}_0 \widetilde{\gamma}_2
%- i \frac{|\chi_3| - |\Omega_3|}{2} \widetilde{\gamma}_0 \widetilde{\gamma}_3
%\end{split}
%\label{Hamiltonian_CaseIII}
%\end{align}
%\begin{align}
%\begin{split}
%H&= -i \frac{|\chi_1| + |\Omega_1|}{2} \gamma_0 \gamma_1
%- i \frac{|\chi_1| - |\Omega_1|}{2} \widetilde{\gamma}_0 \widetilde{\gamma}_1 \\
%&- i \cos\varphi_2 \frac{|\chi_2| + |\Omega_2|}{2} \gamma_0 \gamma_2
%- i \cos\varphi_2 \frac{|\chi_2| - |\Omega_2|}{2} \widetilde{\gamma}_0 \widetilde{\gamma}_2 \\
%&- i \cos\varphi_3 \frac{|\chi_3| + |\Omega_3|}{2} \gamma_0 \gamma_3
%- i \cos\varphi_3 \frac{|\chi_3| - |\Omega_3|}{2} \widetilde{\gamma}_0 \widetilde{\gamma}_3 \\
%&+ i \sin\varphi_2 \frac{|\chi_2| + |\Omega_2|}{2} \widetilde{\gamma}_0 \gamma_2
%- i \sin\varphi_2 \frac{|\chi_2| - |\Omega_2|}{2} \gamma_0 \widetilde{\gamma}_2 \\
%&+ i \sin\varphi_3 \frac{|\chi_3| + |\Omega_3|}{2} \widetilde{\gamma}_0 \gamma_3
%- i \sin\varphi_3 \frac{|\chi_3| - |\Omega_3|}{2} \gamma_0 \widetilde{\gamma}_3
%\end{split}
%\label{Hamiltonian_CaseIII}
%\end{align}
\begin{eqnarray}
%\begin{split}
H&=& -i \frac{|\chi_1| + |\Omega_1|}{2} \gamma_0 \gamma_1
- i \frac{|\chi_1| - |\Omega_1|}{2} \widetilde{\gamma}_0 \widetilde{\gamma}_1\nonumber \\
&-& i \cos\varphi_2\bigr(\frac{|\chi_2| + |\Omega_2|}{2} \gamma_0 \gamma_2
+\frac{|\chi_2| - |\Omega_2|}{2} \widetilde{\gamma}_0 \widetilde{\gamma}_2\bigr) \nonumber\\
&-& i \cos\varphi_3 \bigr(\frac{|\chi_3| + |\Omega_3|}{2} \gamma_0 \gamma_3
+ \frac{|\chi_3| - |\Omega_3|}{2} \widetilde{\gamma}_0 \widetilde{\gamma}_3\bigr)\nonumber \\
&+& i \sin\varphi_2 \bigr(\frac{|\chi_2| + |\Omega_2|}{2} \widetilde{\gamma}_0 \gamma_2
- \frac{|\chi_2| - |\Omega_2|}{2} \gamma_0 \widetilde{\gamma}_2\bigr)\nonumber \\
&+& i \sin\varphi_3 \bigr(\frac{|\chi_3| + |\Omega_3|}{2} \widetilde{\gamma}_0 \gamma_3
- \frac{|\chi_3| - |\Omega_3|}{2} \gamma_0 \widetilde{\gamma}_3\bigr).
%\end{split}
\label{Hamiltonian_CaseIII}
\end{eqnarray}
One can notice that for case III in TABLE \ref{table} that $\varphi_2=\varphi_3=0$, the cross-coupling terms like $i\gamma_i \widetilde{\gamma}_j$ vanish. In this way, a unitary symmetry $\mathcal{R}$ defined as $\mathcal{R}_{ij} i\gamma_i \widetilde{\gamma}_j \mathcal{R}_{ij}^{-1} = -i\gamma_i \widetilde{\gamma}_j$ is preserved~\cite{hong2022unitary}, and all these MZMs are separated into two sets as shown in Fig.~\ref{CaseIIIandIV}(a). According to the braiding protocol introduced above, these two independent sets of MZMs $\{ \gamma_i \}$ and $\{ \widetilde{\gamma}_i \}$ are braided simultaneously. It is worth noticing that $\gamma_i$ and $\widetilde{\gamma}_i$ are spatially fully-overlapped. Therefore, the low-lying modes being braided here are actually fully-overlapped Majorana pairs respecting local unitary symmetry $\mathcal{R}$. Then each Majorana pair as a whole is actually complex fermion mode~\cite{Wu_SCPMA} that the corresponding braiding properties can be constructed from the Majorana ones.

In the adiabatic limit $T \gg 1/w, 1/\Delta$ [$w$ is the (average) energy difference between the lowest two-fold states and the second lowest two-fold states, see Fig. \ref{CaseIIIandIV}(c)], the Berry phase for the lowest two-fold states turns out to be the projection of the tensor product of the Majorana one's into the parity-even subspace as $\hat{\mathbb{P}}_{\mathrm{even}}[ \mathrm{diag}\{e^{-i\pi/4}, e^{i\pi/4}\} \otimes \mathrm{diag}\{e^{-i\pi/4}, e^{i\pi/4}\} ] = \mathrm{diag}\{-i, i\}$. Similarly, the Berry phase for the second lowest two-fold states is $\hat{\mathbb{P}}_{\mathrm{even}}[ \mathrm{diag}\{e^{-i\pi/4}, e^{i\pi/4}\} \otimes \mathrm{diag}\{e^{i\pi/4}, e^{-i\pi/4}\} ] = \mathrm{diag}\{1, 1\}$~\cite{SupplementaryMaterials}. Meanwhile, the dynamic phases also contribute%when $T \gg \frac{1}{w}, \frac{1}{\Delta}$
. Hence, the braiding operator reads $U_{\mathrm{case~III}} = e^{i\theta} \left[ e^{-i\delta} \begin{pmatrix}
    -i & 0 \\
    0 & i
\end{pmatrix} \oplus e^{i\delta} \begin{pmatrix}
    1 & 0 \\
    0 & 1
\end{pmatrix} \right]$, where $\theta$ ($\delta$) are the dynamic phases corresponding to the average energy of (energy difference between) the lowest and second lowest two-fold degenerate states.
%In the limit of $1/w \gg T \gg 1/\Delta$ [$w$ denotes the energy difference as shown in Fig. \ref{CaseIIIandIV}(c)], $\theta_1 \approx \theta_2$ so that the dynamic phase can be dropped.

By contrast, for $1/w \gg T \gg 1/\Delta$, the dynamic phase of the four lowest states only contributes a trivial overall phase% and can be dropped
. Moreover, the Berry phase for the Majorana set $\{ \widetilde{\gamma}_j \}$ is an identity $I_{4\times4}$ here since $\widetilde{\gamma}_j $'s are braided in the non-adiabatic limit $1/w \gg T$~\cite{chang2021lecture}. Therefore, in this case, the braiding operator is obtained by projecting $\mathrm{diag}\{e^{-i\pi/4}, e^{i\pi/4}\} \otimes I_{4\times4}$ into the parity-even subspace, giving $U_{\mathrm{case~III}} = \mathrm{diag}\{e^{-i\pi/4}, e^{-i\pi/4}, e^{i\pi/4}, e^{i\pi/4}\}$~\cite{SupplementaryMaterials}.

Finally, for case \uppercase\expandafter{\romannumeral 4} in TABLE \ref{table} that $\varphi_2,\varphi_3\ne0$, the unitary symmetry $\mathcal{R}$ is broken and the two sets of MZMs $\{ \gamma_i \}$ and $\{ \widetilde{\gamma}_i \}$ are coupled via the cross-coupling terms [see Fig. \ref{CaseIIIandIV}(b)]. The energy spectra in Fig. \ref{CaseIIIandIV}(d) shows that all the two-fold degeneracies are still preserved, while the energies of these degenerate states vary during the braiding. Through numerical calculation~\cite{SupplementaryMaterials}, we have confirmed that the Berry phase accumulated in case \uppercase\expandafter{\romannumeral 4} is exactly the same as the one in case \uppercase\expandafter{\romannumeral 3} (up to a gauge transform). This is in analogy with the consequence stated above that the Berry phase is the same for case \uppercase\expandafter{\romannumeral 2} and case \uppercase\expandafter{\romannumeral 1}. %Therefore, the braiding operator in case \uppercase\expandafter{\romannumeral 4} is also in the form of $U_{\mathrm{case~IV}} = e^{i\theta_1} \sigma_0 \oplus e^{i\theta_2}(-i\sigma_z)$~\cite{SupplementaryMaterials},
The difference lies in that now the dynamic phases $\theta\mp\delta$ are $\varphi_2$-, $\varphi_3$-, and $|\Omega_j|/|\chi_j|$-dependent since the energy spectra vary with $\varphi_2$, $\varphi_3$, and $|\Omega_j|/|\chi_j|$. Meanwhile, the braiding results still exhibit an oscillation with increasing $T$~\cite{SupplementaryMaterials}. Moreover, the (average) energy difference $w$ between the two-fold ground states and the second-lowest two-fold states also increases with $\varphi_2$, $\varphi_3$, and $\big| |\chi_j|-|\Omega_j| \big|$. In this way, the $1/w \gg T \gg 1/\Delta$ condition can no longer be satisfied %and the dynamic phase cannot be dropped
when $\varphi_2$, $\varphi_3$, or $\big| |\chi_j|-|\Omega_j| \big|$ is significantly non-zero. Consequently, an additional dynamic phase is always accumulated alongside the geometric phase, a clear signature of the symmetry-breaking effect~\cite{Wu_NSR} on the braiding of Dirac fermion zero modes.
%has been presented in previous works~\cite{Wu_HOTI, Wu_SCPMA, Wu_NSR} on the braiding of Dirac fermion zero modes when the unitary symmetry is broken.

%%%%%%%%%%%%%%%%%%%% I AM THE CUTTING LINE %%%%%%%%%%%%%%%%%%%%

\textit{Discussions.}
The ``sweet spot'' condition has long been a key objective in experiments~\cite{threeSite_KitaevChain_Kouwenhoven, threeSite_KitaevChain, probeMajorana_KitaevChain_Kouwenhoven, Chun-Xiao_tuning}. We have shown here the non-trivial phases with symmetry protected MZM pairs or Dirac zero modes in the regime away from this condition.
With the removal of fine-tuning condition $|\Omega_j| = |\chi_j|$, the minimal Kitaev chain model becomes more realistic for experimental realization. %experimentally feasible, particularly in ultracold atom systems.
Particularly, our prediction is of high experimental feasibility by setting $|\chi_j| > |\Omega_j| = 0$, as the complexity associated with the superconducting pairing is circumvented. %In this case, the low-lying modes are Dirac fermion modes (case \uppercase\expandafter{\romannumeral 3} and case \uppercase\expandafter{\romannumeral 4}), other than MZMs.
In this case, the low-lying modes being braided are Dirac fermion modes that the fermion number is conserved during the braiding. The minimal Kitaev chain has recently been experimentally engineered in QD arrays~\cite{threeSite_KitaevChain_Kouwenhoven, threeSite_KitaevChain, probeMajorana_KitaevChain_Kouwenhoven} and could also be implemented by utilizing the internal degrees of freedom of ultracold atoms. Taking $^{171}$Yb fermions as an example, the four braiding modes correspond to the atomic spin states,
$^1$S$_0$$|F=1/2,m_F=\pm 1/2\rangle$, and $^3P$$_0$$|F=1/2,m_F = \pm 1/2\rangle$, which have significantly long lifetime.
The tunnelings among the four fermion modes are highly controllable and readily realized by manipulating Raman and clock transitions~\cite{Yb_ColdAtom}. This study paves the way for the ultimate realization of non-Abelian statistics in experiment.

%%%%%%%%%%%%%%%%%%%% I AM THE CUTTING LINE %%%%%%%%%%%%%%%%%%%%

\textit{Acknowledgements.}
The authors thank Leo P. Kouwenhoven and Chun-Xiao Liu for fruitful discussions. This work was financially supported by the Innovation Program for Quantum Science and Technology (Grant No. 2021ZD0302400 and No. 2021ZD0302000), the National Natural Science Foundation of China (Grant No. 12304194, No. 12425401 and No. 12261160368), National Key Research and Development Program of China (Grant No. 2021YFA1400900 and No. 2024YFA1409000), Shanghai Municipal Science and Technology (Grant No. 24DP2600100), Shanghai Municipal Science and Technology Major Project (Grant No.2019SHZDZX01), and by Shanghai Science and Technology Innovation Action Plan (Grant No. 24LZ1400800).

\bibliography{reference}

\end{document}